\RequirePackage[T1]{fontenc}
\documentclass[12pt]{article}
\usepackage[height=8.85in,width=6.45in]{geometry}

\usepackage[utf8]{inputenc}

\usepackage{amsmath}
\usepackage{amssymb}
\usepackage{mathtools}
\usepackage{amsthm}

\usepackage{times}
\usepackage[scaled]{couriers}
\usepackage{mathrsfs}

\usepackage{graphicx}
\usepackage{subcaption}

\usepackage{tikz}
\usepackage{tikz-cd}
\usetikzlibrary{calc}

\usepackage[svgnames]{xcolor}
\usepackage[colorlinks,linktocpage=true,citecolor=DarkGreen,linkcolor=FireBrick]{hyperref}
\usepackage{cite}

\usepackage{bm}
\usepackage{slashed}
\usepackage{braket}

\usepackage{extarrows}

\numberwithin{equation}{section}

\theoremstyle{plain}
\newtheorem{thm}{Theorem}

\theoremstyle{definition}

\numberwithin{thm}{section}

\def\d{{\rm d}}
\def\i{{\mathsf i}}

\def\Im{\mathop{\mathrm{Im}}}

\def\cO{{\cal O}}

\def\cQ{{\cal Q}}

\def\bC{{\mathbb C}}

\def\bP{{\mathbb P}}

\def\bR{{\mathbb R}}

\def\bZ{{\mathbb Z}}

\def\sM{{\mathsf M}}

\def\sS{{\mathsf S}}
\def\sT{{\mathsf T}}

\def\sX{{\mathsf X}}

\def\U{\mathrm{U}}

\def\Spin{\mathrm{Spin}}

\def\SL{\mathrm{SL}}
\def\Mp{\mathrm{Mp}}

\usepackage{amsmath}
\usepackage{mathtools}

\def\del{\partial}
\def\slope{\alpha'}
\def\taubar{\overline{\tau}}
\def\meff{m_{\text{eff}}}

\def\zbar{\overline{z}}
\DeclareMathOperator{\ord}{ord}

\begin{document}

\begin{titlepage}

\begin{flushright}
OU-HET-1321\\
YITP-26-109\\
TU-1316 
\end{flushright}

\vskip 3cm

\begin{center}

{\large \bfseries 7-branes and $\Gamma_0(2)$ Duality in Type 0B String Theory}

\vskip 1cm
Naoto~Kan$^1$, Masashi~Kawahira$^2$, and Hiroki~Wada$^3$
\vskip 1cm

\begin{tabular}{ll}
$^1$ &Department of Physics, The University of Osaka,
Toyonaka 560-0043, Japan \\
$^2$ & Yukawa Institute for Theoretical Physics, Kyoto University, Kyoto 606-8502, Japan\\
$^3$ & Department of Physics, Tohoku University, Sendai 980-8578, Japan\\
\end{tabular}

\vskip 1cm

\end{center}

\noindent
We investigate 7-branes in type 0B string theory, motivated by the recent proposal of Baykara, Dudas, and Vafa, which connects type 0B string theory with M-theory.
On the $Q$-symmetric branch, the self-duality group of type 0B string theory is $\Gamma_0(2)$, whose Abelianization is $\mathbb{Z}\times\mathbb{Z}_4$. 
This implies the existence of two independent 7-brane charges. 
We construct explicit charge operators in terms of modular forms and identify the corresponding charged objects in gravity solutions. 
One is the ordinary D7-brane, while the other is a 7-brane carrying a nontrivial $\mathbb{Z}_4$ charge. 
We further construct a compact background consisting of eight such $\mathbb{Z}_4$ 7-branes using a restricted Weierstrass model. 
Remarkably, in this background the axio-dilaton is fixed to the constant value $\tau=(1+\i)/2$, which coincides with the unstable de Sitter critical point proposed in the M-theory description of type 0B string theory. 
We also analyze fluctuations in the $Q$-odd sector and find that, although the corresponding equations of motion are globally well-defined in this background, the closed string tachyon remains unstable. 
Our results provide a concrete realization of new 7-branes in type 0B string theory and clarify their relation to its strong-coupling structure.

\end{titlepage}

\setcounter{tocdepth}{3}


\tableofcontents

\newpage

\section{Introduction and Summary}

\subsection{Overview}
 
Superstring theory is a UV-complete framework that can, in principle,
describe physics all the way up to the Planck scale, and it is natural
to ask whether it can accommodate the universe we observe.
Observations of the cosmic microwave background exhibit a nearly
scale-invariant primordial spectrum~\cite{Planck:2018vyg}, which is
naturally explained by an early stage of quasi-de~Sitter expansion.
Nevertheless, controlled realizations of de~Sitter spacetime in string
theory remain notoriously
difficult~\cite{Maldacena:2000mw,Obied:2018sgi}.
 
Since de~Sitter backgrounds do not admit unbroken supersymmetry, any
stringy realization of them necessarily involves supersymmetry
breaking.  It is therefore natural to explore intrinsically
non-supersymmetric string
theories~\cite{Dixon:1986iz,Seiberg:1986by,Alvarez-Gaume:1986ghj}, which may offer
qualitatively different possibilities for backgrounds with positive
vacuum energy.  Without supersymmetry, however, most of the standard
tools are unavailable, and the strong-coupling structure of these
theories is poorly understood.
 
A classic step in this direction is the work of Bergman and
Gaberdiel~\cite{Bergman:1999km}, which discussed possible relations
between M-theory~\cite{Witten:1995ex} and type~0 string theories~\cite{Dixon:1986iz}.  Recently, Baykara,
Dudas, and Vafa revisited this relation~\cite{Baykara:2026gem} and
proposed two M-theoretic realizations of type~0B string theory: one is
the compactification of M-theory on $S^1\times S^1$ with antiperiodic
boundary conditions for fermions along one of the circles~\cite{Scherk:1979zr}, and the
other is the compactification on $(S^1\vee S^1)\times S^1$.\footnote{
There are recent works related to these proposals \cite{Altavista:2026brr,Altavista:2026evd,Altavista:2026muc,Altavista:2026edv,Baykara:2026jzs,Baykara:2026vdc,Basile:2026trt,Kamal:2026msr,Fraiman:2026ltu,Garousi:2026nvz,Dasgupta:2026maq}.
}  
In this paper, we mainly adopt the first description, which is directly
connected to the established chain of dualities through type~IIA
superstring theory~\cite{Imamura:1999um}.  Notably, this framework has been
argued to admit an unstable de~Sitter critical point at strong
coupling~\cite{Baykara:2026gem}, which provides an additional
motivation to understand its nonperturbative structure.
 
The purpose of this paper is to investigate which extended objects
exist in type~0B string theory once the strong-coupling regime is
taken into account.  Our guide is the self-duality group.  On the
$Q$-symmetric branch, where all fields odd under the left-moving
fermion parity $Q=(-1)^{G_{\rm L}}$ are turned off, the self-duality
group of type~0B string theory is the congruence subgroup
$\Gamma_0(2)\subset\SL(2,\bZ)$~\cite{Baykara:2026gem}.  In type~IIB
superstring theory, the $\SL(2,\bZ)$ self-duality~\cite{Hull:1994ys} underlies the
existence of 7-branes~\cite{Vafa:1996xn,Greene:1989ya}; in the same
spirit, we propose that 7-branes exist in type~0B string theory, with
monodromies valued in $\Gamma_0(2)$.

Our results are summarized as follows.  First, we formulate the
universal charge of 7-branes as 
the Abelianization map of the duality group.  In type~IIB superstring
theory, this reproduces the familiar $\bZ_{12}$-valued count of
7-branes~\cite{DeWolfe:1998pr}, and an explicit charge formula in terms of the modular
function $\Delta(\tau)=\eta(\tau)^{24}$ was given in
Ref.~\cite{Kan:2025lfb} under the assumption that the axio-dilaton is
a holomorphic function.  We show that this assumption can be removed:
the formula remains valid without holomorphy of the axio-dilaton, and
hence in backgrounds without supersymmetry.  In type~0B string theory,
the Abelianization
\begin{align}
\Gamma_0(2)^{\rm ab}\cong\bZ\times\bZ_4
\end{align}
implies the existence of two independent universal charges, and we
construct the corresponding charge operators explicitly in terms of
the modular forms $\eta(2\tau)^{24}/\eta(\tau)^{24}$ and
$\eta(2\tau)^{8}/\eta(\tau)^{16}$.

Second, we identify the charged objects in gravity solutions.  Under
the $Q$-symmetric ansatz, the low-energy effective action of type~0B
string theory coincides with that of type~IIB supergravity, so that
F-theoretic techniques apply, with the Weierstrass model restricted so
that all monodromies lie in $\Gamma_0(2)$.  The object charged under
the $\bZ$ factor is the ordinary D7-brane, while the generator of the
$\bZ_4$ factor is realized by an exotic 7-brane, which we call the
$\bZ_4$ 7-brane.  Using the restricted Weierstrass model
$y^2=x^3+b(z)x$, we construct a compact background consisting of eight
$\bZ_4$ 7-branes.  In this background the axio-dilaton is frozen at
\begin{align}
\tau=\frac{1+\i}{2},
\end{align}
the unique value invariant under the generator of the $\bZ_4$ factor
of $\Gamma_0(2)$, and this value precisely coincides with the unstable
de~Sitter critical point proposed in Ref.~\cite{Baykara:2026gem}.
 
Third, we analyze the linearized fluctuations of the $Q$-odd fields,
including the closed string tachyon, around these backgrounds.
Requiring their equations of motion to be globally well-defined
constrains the allowed backgrounds: the requirement is satisfied by
the compact $\bZ_4$ background and by the local D7-brane solution,
whereas it fails for the generic restricted Weierstrass background, at
least as long as the $Q$-odd fields are assumed to be neutral under
$\Gamma_0(2)$.  In the compact $\bZ_4$ background the effective mass
of the tachyon receives no contribution from the RR fields, so that
the flat-space tachyonic instability persists: the $\bZ_4$ 7-branes
neither remove nor enhance it.

\subsection{Organization of the paper}

The rest of the paper is organized as follows.

\begin{itemize}

\item 
In Section~\ref{sec:Self_Duality_and_7-brane}, we review the self-dualities in type IIB superstring theory and type 0B string theory.
As is well known, 
type IIB superstring theory is realized by the toroidal compactification of M-theory, and the self-duality group $\SL(2,\bZ)$ arises from the geometric symmetry of the torus.
Recently, 
in Ref.~\cite{Baykara:2026gem} 
the authors argued that type 0B string theory is also realized by the toroidal compactification of M-theory with two kinds of boundary conditions for fermions: 
one is a periodic boundary condition, and the other is an antiperiodic boundary condition.
On the $Q$-symmetric branch, the relevant self-duality group is reduced from $\SL(2,\bZ)$ to $\Gamma_0(2)$.

\item 
In 
Section~\ref{sec:charge}, 
we clarify the notion of the charge of 7-branes.
In Section~\ref{subsec:charge_IIB}, we revisit the case of type IIB superstring theory.
The 7-branes have charges with a $\mathbb{Z}_{12}$ structure, and we find the explicit charge formula.
The novelty is that our formula is valid even in the case without supersymmetry.
In Section~\ref{subsec:charge_0B},
we consider the case of type 0B string theory and find two kinds of charges of 7-branes, unlike the case of type IIB superstring theory.
One has a $\mathbb{Z}$-structure, and the other has a $\mathbb{Z}_4$-structure.
This fact suggests that there are two kinds of 7-branes: ordinary D7-branes with a $\mathbb{Z}$-structure and new exotic 7-branes with a $\mathbb{Z}_4$-structure.
We call the latter $\mathbb{Z}_4$ 7-branes.
In addition, we construct the explicit formulae for the charges of D7-branes and $\mathbb{Z}_4$ 7-branes.

\item
In Section~\ref{sec:7brane_sol},
we construct 7-brane solutions in type 0B gravity theory. We first review 7-brane solutions in type IIB supergravity and their description in terms of the Weierstrass model. We then show that, under the $Q$-symmetric ansatz, analogous solutions exist in type 0B gravity theory. Using a restricted Weierstrass model with $\Gamma_0(2)$ monodromies, we identify gravity solutions carrying the two types of 7-brane charges found in Section \ref{sec:charge}. We also examine the linearized equations of motion for the $Q$-odd fields and discuss the conditions under which these backgrounds can be consistently defined beyond the $Q$-symmetric truncation.

\item
Section~\ref{sec:Discussions} is devoted to discussions.
We comment on the relation between the constant-axio-dilaton background constructed in Section~\ref{subsec:0B_7brane_sol} and the unstable de Sitter solution proposed in Ref.~\cite{Baykara:2026gem}. We also discuss the possible $\Gamma_0(2)$ transformation properties of the $Q$-odd fields, the degrees of freedom living on the $\mathbb{Z}_4$ 7-branes, and a possible extension of our analysis to the conjugate duality group $\Gamma^0(2)$.

\item
Finally, Appendix~\ref{sec:Charge_Weierstrass} explains the relation between the charges of 7-branes and the restricted Weierstrass model.
Using the modular $\lambda$-function and the roots of the cubic equation defining the elliptic fiber, we determine the monodromies around its singular fibers and identify the corresponding $\mathbb{Z}$ and $\mathbb{Z}_4$ charges used in Section~\ref{sec:7brane_sol}.

\end{itemize}

\section{Review of Self-Duality}
\label{sec:Self_Duality_and_7-brane}

\subsection{Self-Duality in Type IIB Superstring Theory}

Let us first recall the origin of the self-duality group of type IIB superstring theory. It arises from the toroidal compactification of M-theory:
\begin{align}
{\rm M\ theory\ on\ }
S^1(R_{10}) \times S^1(R_{11})
\
\xlongrightarrow{R_{10},R_{11}\ll \sqrt{\alpha'}}
\
{\rm IIB\ theory}.
\end{align}
Here, the tenth and eleventh dimensions are compactified as
\begin{align}
x_{10}+R_{10}\sim x_{10},
\ \ 
x_{11}+R_{11}\sim x_{11}.
\end{align}
The complex modulus $\tau$ of the torus is identified with the axio-dilaton of type IIB superstring theory, 
which is the combination of the Ramond-Ramond 0-form $C^{(0)}$ and the dilaton $\phi$ as
\begin{align}
\tau
=C^{(0)}+\i e^{-\phi}.
\end{align}
Moreover, the symmetry $\SL(2,\bZ)$ of the torus becomes its self-duality group.

\subsection{Self-Duality in Type 0B String Theory}
\label{subsec:Self_Duality_0B}

We now determine the self-duality group of type 0B string theory in an analogous manner. 
The discussion in this subsection is based on Ref.~\cite{Baykara:2026gem}.
The key ingredient is the following T-duality~\cite{Imamura:1999um}:
\begin{align}
{\rm IIA\ theory\ on\ }
S^1_{\rm A}(R_{10})
=
{\rm 0B\ theory\ on\ }
S^1_{Q}(2\alpha'/R_{10}).
\label{eq:IIA_0B_T_duality}
\end{align}
Here, the theory on the left-hand side is orbifolded by the spacetime fermion parity $(-1)^{F}$, whereas the theory on the right-hand side is orbifolded by the left-moving fermion parity $Q=(-1)^{G_{\rm L}}$.\footnote{
This is the quantum symmetry\cite{Vafa:1989ih} arising from orbifolding.
}
It therefore follows that
\begin{align}
{\rm M\ theory\ on\ }
S^1_{\rm A}(R_{10})\times S^1_{\rm P}(R_{11})
\
\xlongrightarrow{R_{10},R_{11}\ll \sqrt{\alpha'}}
\
{\rm 0B\ theory}.
\end{align}
Here, $S^1_{\rm P}$ denotes a circle without orbifolding. 
Thus, in the case of type 0B string theory, we consider the symmetry of $T^2_{\rm AP}:=S^1_{\rm A}\times S^1_{\rm P}$, rather than the usual torus symmetry $\SL(2,\bZ)$.
We can see $T^2_{\rm AP}$ as a torus with a $(-1)^F$ topological line.
\begin{align}
\begin{tikzpicture}[baseline=(base.center)]
  \node[inner sep=0pt] (base) at (1.5,1.5) {};
  \draw[thick] (0,0) rectangle (3,3);
  \node[left=2pt] at (0,1.5) {${\rm A}$};
  \node[below=2pt] at (1.5,0) {${\rm P}$};
\end{tikzpicture}
\ \  = \ \ 
\begin{tikzpicture}[baseline=(base.center)]
  \node[inner sep=0pt] (base) at (1.5,1.5) {};
  \draw[thick] (0,0) rectangle (3,3);
  \draw[red,thick] (0,1.5) -- (3,1.5);
  \node[above=2pt,red] at (1.5,1.5) {$(-1)^F$};
\end{tikzpicture}
\end{align}
Under the $\sT$-transformation, the torus $T^2_{\rm AP}$ is invariant.
\begin{align}
\sT:
\begin{tikzpicture}[baseline=(base.center)]
  \node[inner sep=0pt] (base) at (1.5,1.5) {};
  \draw[thick] (0,0) rectangle (3,3);
  \draw[red,thick] (0,1.5) -- (3,1.5);
\end{tikzpicture}
\ \ \mapsto\ \ 
\begin{tikzpicture}[baseline=(base.center)]
  \node[inner sep=0pt] (base) at (1.5,1.5) {};
  \draw[thick] (0,0) rectangle (3,3);
  \draw[red,thick] (0,1.5) -- (3,1.5);
\end{tikzpicture}
\end{align}
However, under the $\sS$-transformation, it is not invariant.
\begin{align}
\sS:
\begin{tikzpicture}[baseline=(base.center)]
  \node[inner sep=0pt] (base) at (1.5,1.5) {};
  \draw[thick] (0,0) rectangle (3,3);
  \draw[red,thick] (0,1.5) -- (3,1.5);
\end{tikzpicture}
\ \ \mapsto\ \ 
\begin{tikzpicture}[baseline=(base.center)]
  \node[inner sep=0pt] (base) at (1.5,1.5) {};
  \draw[thick] (0,0) rectangle (3,3);
  \draw[red,thick] (1.5,0) -- (1.5,3);
\end{tikzpicture}
\end{align}
We define the $\sX$ matrix by
\begin{align}
\sX
:=
(\sT\sS)^{-1}\sS(\sT\sS)
=
\begin{pmatrix}
 1&-1\\
 2&-1
\end{pmatrix}.
\end{align}
Interestingly, under the $\sX$-transformation, $T^2_{\rm AP}$ is invariant.
\begin{align}
\begin{tikzpicture}[baseline=(base.center)]
  \node[inner sep=0pt] (base) at (1.5,1.5) {};
  \draw[thick] (0,0) rectangle (3,3);
  \draw[red,thick] (0,1.5) -- (3,1.5);
\end{tikzpicture}
&\ \ \overset{\sS}{\mapsto}\ \ 
\begin{tikzpicture}[baseline=(base.center)]
  \node[inner sep=0pt] (base) at (1.5,1.5) {};
  \draw[thick] (0,0) rectangle (3,3);
  \draw[red,thick] (1.5,0) -- (1.5,3);
\end{tikzpicture}
\ \ \overset{\sT}{\mapsto}\ \ 
\begin{tikzpicture}[baseline=(base.center)]
  \node[inner sep=0pt] (base) at (1.5,1.5) {};
  \draw[thick] (0,0) rectangle (3,3);
  \draw[red,thick] (1.5,0) -- (3,1.5);
  \draw[red,thick] (0,1.5) -- (1.5,3);
\end{tikzpicture}
\notag
\\
&\ \ \overset{\sS}{\mapsto}\ \ 
\begin{tikzpicture}[baseline=(base.center)]
  \node[inner sep=0pt] (base) at (1.5,1.5) {};
  \draw[thick] (0,0) rectangle (3,3);
  \draw[red,thick] (1.5,3) -- (3,1.5);
  \draw[red,thick] (0,1.5) -- (1.5,0);
\end{tikzpicture}
\ \ \overset{}{=}\ \ 
\begin{tikzpicture}[baseline=(base.center)]
  \node[inner sep=0pt] (base) at (1.5,1.5) {};
  \draw[thick] (0,0) rectangle (3,3);
  \draw[red,thick] (1.5,0) -- (3,1.5);
  \draw[red,thick] (0,1.5) -- (1.5,3);
\end{tikzpicture}
\notag
\\
&\ \ \overset{\sT^{-1}}{\mapsto}\ \ 
\begin{tikzpicture}[baseline=(base.center)]
  \node[inner sep=0pt] (base) at (1.5,1.5) {};
  \draw[thick] (0,0) rectangle (3,3);
  \draw[red,thick] (1.5,0) -- (1.5,3);
\end{tikzpicture}
\ \ \overset{\sS^{-1}}{\mapsto}\ \ 
\begin{tikzpicture}[baseline=(base.center)]
  \node[inner sep=0pt] (base) at (1.5,1.5) {};
  \draw[thick] (0,0) rectangle (3,3);
  \draw[red,thick] (0,1.5) -- (3,1.5);
\end{tikzpicture}
\end{align}
Actually, the symmetry of $T^2_{\rm AP}$ is known as
\begin{align}
\Gamma_0(2)
:=&
\left\{
\begin{pmatrix}
a&b\\
c&d
\end{pmatrix}
\in\SL(2,\bZ)
\ \middle|\ 
c\equiv 0\ ({\rm mod}\ 2)
\right\}
\notag\\
=&
\langle 
\sT,\sX
\ |\ 
\sX^2=-1
\rangle.
\end{align}

Consider a state of the right-hand side in Eq.~\eqref{eq:IIA_0B_T_duality}
\begin{align}
|Q,n\rangle
\end{align}
whose Kaluza--Klein momentum along the $x^{10}$-direction is given by $p=n/R_{10}$. This state satisfies
\begin{align}
Q=(-1)^n
\end{align}
and hence the zero mode is $Q$-even.
In other words, $Q$-odd states in type 0B string theory correspond to winding modes in type IIA superstring theory.
In M-theory, they are not geometrized and are instead realized as wrapped M2-brane modes.

The closed string spectrum of type 0B string theory is given by the following table.
\begin{align}
\begin{tabular}{|c|c|}
\hline
{\rm sector}
&
{\rm field}
\\
\hline
$({\rm NS}^+,{\rm NS}^+)$
&
$g_{\mu\nu},B_{\mu\nu},\phi$
\\
$({\rm NS}^-,{\rm NS}^-)$
&
$T$
\\
$({\rm R}^+,{\rm R}^+)$
&
$
C^{(0)},
C^{(2)}_{\mu\nu},
C^{(4)}_{\mu\nu\rho\sigma}
$
\\
$({\rm R}^-,{\rm R}^-)$
&
$
\tilde{C}^{(0)},
\tilde{C}^{(2)}_{\mu\nu},
\tilde{C}^{(4)}_{\mu\nu\rho\sigma}
$
\\
\hline
\end{tabular}
\end{align}
Therefore, the $Q$ eigenvalues of the fields are as follows.
\begin{align}
\begin{tabular}{|c|c|}
\hline
$Q$
&
{\rm field}
\\
\hline
$+$
&
$
g_{\mu\nu},
B_{\mu\nu},
\phi,
C^{(0)},
C^{(2)}_{\mu\nu},
C^{(4)}_{\mu\nu\rho\sigma}
$
\\
\hline
$-$
&
$
T,
\tilde{C}^{(0)},
\tilde{C}^{(2)}_{\mu\nu},
\tilde{C}^{(4)}_{\mu\nu\rho\sigma}
$
\\
\hline
\end{tabular}
\end{align}
In view of this, the geometric feature in M-theory is encoded in the $Q$-even sector in type 0B string theory.
For example, the complex modulus $\tau$ of the torus $T_{\rm AP}$ is represented as
\begin{align}
\tau=C^{(0)}+\i e^{-\phi},
\label{eq:axio-dilaton_0B}
\end{align}
which can be shown by the relation \eqref{eq:IIA_0B_T_duality}.

Based on the above discussion, we expect that, in type 0B string theory, when all fields in the $Q$-odd sector are set to zero, there is a symmetry acting on the axio-dilaton $\tau$ by $\Gamma_0(2)$ transformations. A notable feature of this group is that it does not contain the $\sS$ transformation:
\begin{align}
\sS\notin \Gamma_0(2).
\end{align}
In other words, this self-duality is not an ordinary strong-weak duality.

It should be noted, however, that this analysis is restricted to the $Q$-even sector, and that this picture is less reliable when fields in the $Q$-odd sector, such as the tachyon $T$, are nonzero.\footnote{In Ref.~\cite{Baykara:2026gem}, it is conjectured that, once the $Q$-odd sector is included, this symmetry is enhanced to $\SL(2,\bZ)$.} Nevertheless, as we discuss in Section~\ref{sec:7brane_sol}, the corresponding gravitational solutions can be constructed explicitly. Moreover, they can be described through a Weierstrass model, as in conventional F-theory.

\section{Charges of 7-branes}
\label{sec:charge}

\subsection{Charges of 7-branes in Type IIB Superstring Theory}
\label{subsec:charge_IIB}

Type IIB superstring theory contains a $(1+7)$-dimensional soliton known as the 7-brane. 
We first revisit the notion of the charge of 7-branes.
In the standard discussion, it is assumed that the background is supersymmetric, i.e., the axio-dilaton $\tau$ is a holomorphic function.
However, our discussion is valid even in backgrounds without supersymmetry.

Let us begin with the simplest case, that of a D7-brane.
The charge of a D7-brane is measured using a closed loop $\gamma$ encircling the brane~\cite{Polchinski:1995mt}.
More precisely, in terms of the Ramond--Ramond 0-form $C^{(0)}$, it is given by
\begin{align}
Q_\gamma=\oint_\gamma {\rm d}C^{(0)}.
\end{align}
Equivalently, a monodromy of $C^{(0)}$ is defined around the D7-brane:
\begin{align}
C^{(0)}\mapsto C^{(0)}+n,
\end{align}
where $n=Q_\gamma$.

A 7-brane is a generalization of a D7-brane.
In the presence of a 7-brane, an $\SL(2,\bZ)$-valued monodromy\footnote{
This originates from the $\SL(2,\bZ)$ symmetry of type IIB superstring theory.
} 
is defined around the brane:
\begin{align}
\tau\mapsto \frac{a\tau+b}{c\tau+d},
\end{align}
where
$
\sM=
\begin{pmatrix}
a&b\\
c&d
\end{pmatrix}
\in\SL(2,\bZ).
$
By taking
$
\sM=
\begin{pmatrix}
1&n\\
0&1
\end{pmatrix},
$
one recovers the monodromy of a D7-brane. Thus, a 7-brane can be regarded as a generalization of a D7-brane.

Let us now consider how the charge of a 7-brane should be defined. We denote by $Q_\gamma(\sM)$ the charge associated with the case in which the monodromy of the 7-branes enclosed by $\gamma$ is $\sM$.
Suppose that there are two 7-branes with monodromies $\sM_1$ and $\sM_2$, respectively. Their total monodromy is then $\sM_2\sM_1$.\footnote{Depending on the orientation of $\gamma$, the total monodromy may instead be $\sM_1\sM_2$, but this distinction does not make a significant difference in the present discussion.}
Let us require the charge $Q_\gamma$ to be additive, namely,
\begin{align}
Q_\gamma(\sM_2\sM_1)=Q_\gamma(\sM_2)+Q_\gamma(\sM_1).
\end{align}
In other words, $Q_\gamma$ is a group homomorphism from $\SL(2,\bZ)$ to an Abelian group $A$.

Here, we recall the following mathematical theorem.

\par\medskip
\noindent
\vrule width 1pt
\hspace{0.8em}
\begin{minipage}[t]{0.94\linewidth}
\begin{thm}
Let $G$ be an arbitrary group and $A$ be an Abelian group, and consider a homomorphism $\phi:G\to A$. 
Let $G^{\rm ab}$ denote the Abelianization of $G$. 
Then there exists a homomorphism $\varphi:G^{\rm ab}\to A$ such that the following diagram commutes:
\[
\begin{tikzcd}
G \arrow[rr,"\phi"] \arrow[dr,"{\rm proj}"'] && A \\
& G^{\rm ab} \arrow[ur,"\varphi"'] &
\end{tikzcd}
\]
Here, ${\rm proj}:G\to G^{\rm ab}$ denotes the canonical projection map defined by $G^{\rm ab}:=G/[G,G]$ where $[G,G]$ denotes the commutator subgroup.
\end{thm}
\end{minipage}
\par\medskip
\

By this theorem, any charge operator $Q_\gamma:\SL(2,\bZ)\to A$ can be regarded as the composition of the universal charge operator
\begin{align}
\mathcal{Q}_\gamma:
\SL(2,\bZ)\to
\SL(2,\bZ)^{\rm ab}
\cong
\bZ_{12}
\end{align}
and a homomorphism
\begin{align}
\varphi:
\bZ_{12}
\to
A.
\end{align}

In fact, the number of 7-branes discussed in conventional F-theory is precisely the universal charge $\mathcal{Q}_\gamma$.\footnote{
The details are discussed in Section~\ref{subsec:IIB_7brane_sol}.
}
Let $\sT$ and $\sS$ be matrices given by
\begin{align}
\sT
:=
\begin{pmatrix}
1&1\\
&1
\end{pmatrix}
,\ \ 
\sS
:=
\begin{pmatrix}
&-1\\
1&
\end{pmatrix},
\end{align}
which are generators of $\SL(2,\bZ)$.
Then $\mathcal{Q}_\gamma(\sT)=1$ and $\mathcal{Q}_\gamma(\sS)=9$ modulo 12.
Hereafter, whenever we refer simply to the charge of a 7-brane, we mean this universal charge.

Here we examine the explicit form of the universal charge operator $\mathcal{Q}_\gamma$.
The universal charge of the 7-branes enclosed by $\gamma$ is given as follows:\footnote{
Although this expression is given in Ref.~\cite{Kan:2025lfb}, 
the authors argue that this expression applies only when $\tau$ is holomorphic, i.e., in a supersymmetric background.
However, as we show below, this is valid without the holomorphic ansatz or supersymmetry.
}\footnote{
If the low-energy physics can be described by quantum field theories, the operator $e^{\i \theta\mathcal{Q}_\gamma}$ can be seen as a topological operator, and there exists a generalized symmetry\cite{Kapustin:2014gua,Gaiotto:2014kfa}, whose group structure is $\mathbb{Z}_{12}$. 
If one considers the fermionic sector, the symmetry enhances to $\mathbb{Z}_{24}$ \cite{Kan:2025lfb}.
This is consistent with the cobordism conjecture.
As explained in Refs.~\cite{Dierigl:2022reg,McNamara:2021cuo},
the bordism group $\Omega^{\Spin-G}_1({\rm pt})$ is given by $G^{\rm ab}$, and
\begin{align}
\Omega^{\Spin-\Mp(2,\bZ)}_1({\rm pt})
\cong
\bZ_{24},
\notag
\end{align}
which represents 7-branes.
}
\begin{align}
\mathcal{Q}_\gamma
=
\frac{1}{2\pi \i}
\oint_\gamma
{\rm d}x^\mu
\frac{{\rm d}}{{\rm d}x^\mu}
\log\Delta(\tau(x)).
\label{eq:universal_charge_in_IIB}
\end{align}
Here, $\Delta(\tau)$ is defined by
\begin{align}\label{eq:discriminant_eta_IIB}
\Delta(\tau):=\eta(\tau)^{24}.
\end{align}

However, the one-form
\begin{align}
{\rm d}x^\mu\frac{{\rm d}}{{\rm d}x^\mu}\log\Delta(\tau(x))
\end{align}
is not globally well-defined.
Therefore, more precisely, the closed curve $\gamma$ must be divided into segments
\begin{align}
\gamma_1\cup\gamma_2\cup\cdots\cup\gamma_n
\end{align}
and the transition functions must also be taken into account.
Let $\gamma_i$ be a path connecting $x_{i-1}$ to $x_{i}$, with $x_{n}=x_{0}$.
Suppose that the transition function from $\gamma_i$ to $\gamma_{i+1}$ is given by
\begin{align}
\tau(x_{i})|_{\gamma_{i+1}}
=
\sM_i\cdot\tau(x_{i})|_{\gamma_i}
,
\qquad
\sM_i=
\begin{pmatrix}
a_i&b_i\\
c_i&d_i
\end{pmatrix}
\in\SL(2,\bZ).
\end{align}
Then, the change induced by the transition function is
\begin{align}
\log\Delta(
\tau(x_{i})|_{\gamma_{i+1}}
)
-
\log\Delta(
\tau(x_{i})|_{\gamma_{i}}
)
=
12\log(c_i\tau(x_{i})|_{\gamma_i}+d_i).
\end{align}
Thus, the precise expression is given by
\begin{align}
&\mathcal{Q}_\gamma
=
\frac{1}{2\pi\i}
\sum_{i=1}^n
\left[
\int_{\gamma_i}
{\rm d}x^\mu
\frac{{\rm d}}{{\rm d}x^\mu}
\log\Delta(\tau_i(x))
-
12F(\tau_i(x_i),\sM_i)
\right]
\label{eq:precise_charge}
\end{align}
where $\tau_i(x):=\tau(x)|_{\gamma_i}$ and $F(\tau_i,\sM_i):=\log(c_i\tau_i+d_i)$.

The first term in Eq.~\eqref{eq:precise_charge} is independent of the choice of branch, whereas the second term depends on the choice of branch,
\begin{align}
F(\tau_i,\sM_i)
\mapsto
F(\tau_i,\sM_i)
+
2\pi\i n_i,
\ \ 
(n_i\in\mathbb{Z}).
\end{align}
Therefore, the charge $\mathcal{Q}_\gamma$ is well-defined only modulo 12.

In fact, $\mathcal{Q}_\gamma$ is the universal charge.
An explicit computation confirms that
\begin{align}
\mathcal{Q}_\gamma(\sT)=1,
\qquad
\mathcal{Q}_\gamma(\sS)=9
\qquad
(\mathrm{mod}\ 12).
\end{align}
When the monodromy along $\gamma$ is $\sM=\sT$, taking $x_1=x_2$ for the endpoints $x_1$ and $x_2$ of $\gamma_1$, we find
\begin{align}
\mathcal{Q}_\gamma(\sT)
&=
\frac{1}{2\pi\i}
\int_{\gamma_1}
{\rm d}x^\mu
\frac{{\rm d}}{{\rm d}x^\mu}
\log\Delta(\tau(x))
-
12F(\tau(x_1),\sT)
\notag
\\
&=
\frac{1}{2\pi\i}
\left[
\log\Delta(\sT\cdot\tau(x_1))
-
\log\Delta(\tau(x_1))
\right]
\notag
\\
&=
\frac{1}{2\pi\i}
\left[
\log\Delta(\tau(x_1)+1)
-
\log\Delta(\tau(x_1))
\right]
\notag
\\
&=
\frac{1}{2\pi\i}
\times
2\pi\i
\notag
\\
&=
1,
\end{align}
where we used $F(\tau(x_1),\sT)=0$.
Similarly, when $\sM=\sS$, we obtain
\begin{align}
\mathcal{Q}_\gamma(\sS)
&=
\frac{1}{2\pi\i}
\int_{\gamma_1}
{\rm d}x^\mu
\frac{{\rm d}}{{\rm d}x^\mu}
\log\Delta(\tau(x))
-
12F(\tau(x_1),\sS)
\notag
\\
&=
\frac{1}{2\pi\i}
\left[
\log\Delta(\sS\cdot\tau(x_1))
-
\log\Delta(\tau(x_1))
-
12
\log(\tau(x_1))
\right]
\notag
\\
&=
\frac{1}{2\pi\i}
\left[
12\log(-\i\tau(x_1))
-
12
\log(\tau(x_1))
\right]
\notag
\\
&=
\frac{1}{2\pi\i}
\times
12
\times
\left(
-\frac{\pi\i}{2}
\right)
\notag
\\
&=
-3,
\end{align}
where we have chosen the principal value $\log(-\i)=-\pi\i/2$.\footnote{
If the other principal value is chosen, one instead has $\log(-\i)=3\pi\i/2$, which gives $\mathcal{Q}_\gamma=9$.
}

\subsection{Charges of 7-branes in Type 0B String Theory}
\label{subsec:charge_0B}

One can define 7-branes in terms of this $\Gamma_0(2)$ monodromy. Interestingly,
\begin{align}
\Gamma_0(2)^{\rm ab}
\cong
\bZ\times \bZ_4.
\end{align}
This implies that there are two types of universal charge operators:
\begin{align}
\mathcal{Q}^{\bZ}_\gamma:
\Gamma_0(2)
\to
\bZ
,
\qquad
\mathcal{Q}^{\bZ_4}_\gamma:
\Gamma_0(2)
\to
\bZ_4,
\end{align}
and they satisfy
\begin{align}
\begin{cases}
\mathcal{Q}^{\bZ}_\gamma(\sT)=1
,
\qquad
\mathcal{Q}^{\bZ}_\gamma(\sX)=0
,\\
\mathcal{Q}^{\bZ_4}_\gamma(\sT)=0
,
\qquad
\mathcal{Q}^{\bZ_4}_\gamma(\sX)=1.
\end{cases}
\end{align}
Accordingly, this suggests the existence of two types of 7-branes.

We now present explicit expressions for the universal charge operators.
As discussed above, the modular function $\Delta(\tau)$ plays a central role in defining the charge operator in type IIB superstring theory.
To this end, we introduce the following  functions,\footnote{
A similar function 
\begin{align}
\tilde\Delta(\tau)
=
\frac{\eta(2\tau)^{16}}{\eta(\tau)^8}
\end{align}
was originally introduced in Ref.~\cite{tsutsumi2007atkin} as a $\Gamma_0(2)$ analogue of the modular function.
These functions satisfy the relation:
$\Delta^{\bZ}(\tau)/\Delta^{\bZ_4}(\tau)=\tilde\Delta(\tau).$
}
\begin{align}\label{eq:modular_func_0B}
\Delta^{\bZ}(\tau)
:=
\frac{\eta(2\tau)^{24}}{\eta(\tau)^{24}},\ \ 
\Delta^{\bZ_{4}}(\tau)
:=
\frac{\eta(2\tau)^{8}}{\eta(\tau)^{16}}.
\end{align}
We then define the charges
\begin{align}\label{eq:charge_op_0B}
\mathcal{Q}^{\bZ}_\gamma
:=&
\frac{1}{2\pi\i}
\oint_\gamma
\d x^\mu\frac{\d}{\d x^\mu}\log\Delta^{\bZ}(\tau(x)),\\ 
\mathcal{Q}^{\bZ_4}_\gamma
:=&
\frac{1}{2\pi\i}
\oint_\gamma
\d x^\mu\frac{\d}{\d x^\mu}\log\Delta^{\bZ_4}(\tau(x)).
\end{align}

The modular function $\Delta^{\bZ}(\tau)$ has the following properties:
\begin{align}
\log\Delta^{\bZ}(\sT\cdot\tau)
&=
\log\Delta^{\bZ}(\tau)
+
2\pi\i,\\
\log\Delta^{\bZ}(\sX\cdot\tau)
&=
\log\Delta^{\bZ}(\tau).
\end{align}
Therefore, the one-form
\begin{align}
\d x^\mu\frac{\d}{\d x^\mu}\log\Delta^{\bZ}(\tau(x))
\end{align}
is globally well-defined. 
Thus, $\mathcal{Q}^{\bZ}_{\gamma}$ is well-defined. 
We can explicitly verify that
\begin{align}
\mathcal{Q}^{\bZ}_\gamma(\sT)=1,
\ \ 
\mathcal{Q}^{\bZ}_\gamma(\sX)=0.
\end{align}

On the other hand, the modular form $\Delta^{\bZ_4}(\tau)$ has the following properties:
\begin{align}
\log\Delta^{\bZ_4}(\sT\cdot\tau)
&=
\log\Delta^{\bZ_4}(\tau)
,\\
\log\Delta^{\bZ_4}(\sX\cdot\tau)
&=
\log\Delta^{\bZ_4}(\tau)
+2\pi\i
-4\log(2\tau-1).
\end{align}
Therefore, the one-form
\begin{align}
\d x^\mu\frac{\d}{\d x^\mu}\log\Delta^{\bZ_4}(\tau(x))
\end{align}
is not globally well-defined.
In order to make it well-defined, we need to treat it in the same manner as in type~IIB superstring theory.
Thus, the precise expression is given by
\begin{align}
&\mathcal{Q}_\gamma
=
\frac{1}{2\pi\i}
\sum_{i=1}^n
\left[
\int_{\gamma_i}
{\rm d}x^\mu
\frac{{\rm d}}{{\rm d}x^\mu}
\log\Delta^{\mathbb{Z}_4}(\tau_i(x))
+
4F(\tau_i(x_i),\sM_i)
\right].
\label{eq:precise_Z4_charge}
\end{align}
$\mathcal{Q}^{\bZ_4}_\gamma$ is well-defined modulo 4.
We can explicitly verify that
\begin{align}
\mathcal{Q}^{\bZ_4}_\gamma(\sT)=0,
\ \ 
\mathcal{Q}^{\bZ_4}_\gamma(\sX)=1
\ \ 
({\rm mod}\ 4).
\end{align}

\section{Gravity Solution}
\label{sec:7brane_sol}

\subsection{7-brane Solutions in Type~IIB Supergravity}
\label{subsec:IIB_7brane_sol}
Before constructing $7$-brane solutions in type~0B gravity theory, we first briefly review $7$-brane solutions in type~IIB supergravity.\footnote{For details of the review presented here, see, e.g., Refs.~\cite{Polchinski:1998rr,Johnson:2003gi}.}
To make our discussion concrete, we focus on supersymmetric backgrounds where $7$-branes extend along $x^{0},x^{1},\dots,x^{7}$ and are localized at points in the $x^{8}$-$x^{9}$ plane.
In such a background, it is convenient to introduce the complex coordinate $z=x^{8}+\i x^{9}$.
With the ansatz that the NSNS $2$-form field and the RR $2$- and $4$-form fields vanish, the preservation of supersymmetry requires that the axio-dilaton $\tau$ be a holomorphic function of the complex coordinate $z$.
For a holomorphic function $\tau(z)$, the equation of motion for the axio-dilaton field is automatically satisfied.
Since the spacetime is assumed to be a direct product of a $(1+7)$-dimensional Minkowski space and the $z$-plane transverse to the $7$-branes, the metric takes the following form,
\begin{align}\label{eq:metric_ansatz}
	\d s^{2}=\eta_{\mu\nu}\d x^{\mu}\d x^{\nu}+e^{\rho(z,\zbar)}\d z\d \zbar.
\end{align}

As a simple example, let us consider a solution for a single D$7$-brane with a monodromy $\sT\in\SL(2,\bZ)$ located at $z=0$.
The condition of the monodromy requires that the axio-dilaton field behaves near $z=0$ as
\begin{align}\label{eq:D7_tau_sol}
	\tau(z)\sim\frac{1}{2\pi\i}\log z+(\text{regular terms}).
\end{align}
Under our ansatz, the only non-trivial equation of motion is the Einstein equation, and the solution is given by
\begin{align}\label{eq:D7_metric_sol}
	e^{\rho(z,\zbar)}=\tau_{2}(z,\zbar)|\eta(\tau(z))|^{4}\left|z\right|^{-1/6},
\end{align}
where $\tau=\tau_{1}+\i\tau_{2}$, and $\eta(\tau)$ is the Dedekind eta function.
Note that due to the deficit angle $2\pi/12$ of the D$7$-brane, the $z$-plane cannot be compactified to a two-sphere.

To describe the background of type~IIB supergravity with general BPS $7$-branes, which are characterized by their monodromies, one can use the F-theory framework~\cite{Vafa:1996xn}.
In this framework, the axio-dilaton field $\tau$ is identified with the complex structure of an elliptic curve, and the $7$-brane background is described by an elliptically fibered Calabi--Yau manifold.
For instance, an elliptically fibered K3 surface can be described by the Weierstrass form,
\begin{align}\label{eq:Weierstrass}
	y^{2}=x^{3}+f(z)x+g(z),
\end{align}
where $f(z)$ and $g(z)$ are polynomials of degree $8$ and $12$, respectively.
F-theory on a K3 surface corresponds to a background of type~IIB supergravity on $\bC\bP^{1}$ containing general $7$-branes.
The loci of the $7$-branes are given by the zeros of the discriminant,
\begin{align}
	\Delta_{W}(z)=-16(4f(z)^3+27g(z)^2),
\end{align}
which is a polynomial in $z$.
In the background described by the F-theory framework, the universal charge~\eqref{eq:universal_charge_in_IIB} can be written as\footnote{The discriminant is related to the modular function $\Delta(\tau)$ defined in~\eqref{eq:discriminant_eta_IIB} by $\Delta_{W}=(4\pi/\omega_{1})^{12}\Delta$. Here, $\omega_{1}$ is the period introduced in Appendix~\ref{subsec:lambda}.}
\begin{align}
	\cQ_\gamma=\sum_{z_*:\,\text{locus inside\,}\gamma} \ord_{z=z_{\ast}} \Delta_{W}(z).
\end{align}
From the F-theory perspective, this expression states that the universal charge counts the $7$-branes with multiplicity $\ord_{z=z_{\ast}}\Delta_{W}$, modulo $12$.

\subsection{7-brane Solutions in Type~0B Gravity Theory}
\label{subsec:0B_7brane_sol}
In contrast to type~IIB supergravity, the low-energy effective gravity theory of type~0B string theory does not possess spacetime supersymmetry.
Hence, at first glance, there is no reason to take the axio-dilaton to be holomorphic in type~0B gravity theory.
However, the holomorphic axio-dilaton field still allows us to construct concrete gravity solutions, which confirm the existence of $7$-branes in type~0B gravity theory. 

To see this, we begin by recalling the action of type~0B gravity theory, which can be found in Refs.~\cite{Klebanov:1998yya,Garousi:2003db}.
Since we are interested in $7$-brane solutions, we focus on the metric $g_{\mu\nu}$, the dilaton field $\phi$, the tachyon field $T$, and the two RR $0$-form fields $C^{(0)}$ and $\tilde{C}^{(0)}$.
As mentioned in Eq.~\eqref{eq:axio-dilaton_0B}, we organize the dilaton field $\phi$ and one RR $0$-form field $C^{(0)}$ into the axio-dilaton field even for type~0B gravity theory.
Since the other massless fields, such as the NSNS $2$-form field and the RR $2$- and $4$-form fields, appear only quadratically in the action, they can be consistently set to zero in the following discussion.
Then, the relevant part of the action in the Einstein frame is given by
\begin{align}\begin{aligned}\label{eq:axio_dilaton_frame}
	S=&\frac{1}{2\kappa^{2}}\int\,\d^{10}x\,\sqrt{-g}\left[
	R-\frac{\del_{\mu}\tau\del^{\mu}\taubar}{2(\Im\tau)^{2}}-\frac{1}{2}\del_{\mu}T\del^{\mu}T-\frac{1}{2}m^{2}e^{\phi/2}T^{2}
	\right.\\
	&\left.\quad-\frac{1}{2}e^{2\phi}\left(
	\del_{\mu}\tilde{C}^{(0)}\del^{\mu}\tilde{C}^{(0)}+\frac{1}{4}\left(\del_{\mu}C^{(0)}\del^{\mu}C^{(0)}+\del_{\mu}\tilde{C}^{(0)}\del^{\mu}\tilde{C}^{(0)}\right)T^{2}
	+\frac{1}{\sqrt{2}}\del_{\mu}C^{(0)}\del^{\mu}\tilde{C}^{(0)}T
	\right)\right],
\end{aligned}\end{align}
where $\kappa$ is the gravitational coupling constant and $m^{2}=-2/\alpha'$ is the mass of the tachyon field.\footnote{Since the sphere amplitudes for three and four tachyon vertex operators vanish, the cubic and quartic terms of the tachyon field are absent at this level~\cite{Klebanov:1998yya,Garousi:2003db}.}

Note that we can consistently turn off the tachyon field $T$ and the second RR $0$-form field $\tilde{C}^{(0)}$ in type~0B gravity theory.
To see this, we check that the truncation is consistent with the equation of motion for $T$, which is given by
\begin{align}\label{eq:eom_tachyon}
	\left(\nabla_{\mu}\nabla^{\mu}-\meff^{2}e^{\phi/2}\right)T=\frac{1}{2\sqrt{2}}e^{2\phi}\del_{\mu}C^{(0)}\del^{\mu}\tilde{C}^{(0)},
\end{align}
where the effective mass $\meff$ is defined as
\begin{align}\label{eq:effective_mass}
	\meff^{2}=m^{2}+\frac{1}{4}e^{3\phi/2}\left(\del_{\mu}C^{(0)}\del^{\mu}C^{(0)}+\del_{\mu}\tilde{C}^{(0)}\del^{\mu}\tilde{C}^{(0)}\right).
\end{align}
This equation of motion is solved by $T=0$ and $\tilde{C}^{(0)}=0$.
The equation of motion for the RR $0$-form field $\tilde{C}^{(0)}$ is likewise consistent with this ansatz.
We refer to this ansatz as the $Q$-symmetric ansatz.

We now comment on the symmetry of type~0B gravity theory.
Under the $Q$-symmetric ansatz, the action for type~0B gravity theory is identical to that of type~IIB supergravity.\footnote{More precisely, the sphere amplitudes of type~0B string theory involving only the vertex operators in the (NS$^{+}$, NS$^{+}$) and (R$^{+}$, R$^{+}$) sectors coincide with those of the type~IIB superstring theory~\cite{Klebanov:1998yya}.}
This implies that, as in type~IIB supergravity, type~0B gravity theory under the $Q$-symmetric ansatz possesses an $\SL(2,\bR)$ symmetry.
The actual symmetry of type~0B string theory is determined by taking into account the full quantum effects.
The argument in Section~\ref{subsec:Self_Duality_0B} suggests that the symmetry group of type~0B string theory is $\Gamma_{0}(2)$ as long as the theory is restricted to the $Q$-symmetric branch.

The coincidence between the actions of type~IIB supergravity and type~0B gravity theory under the $Q$-symmetric ansatz leads to a method for constructing $7$-brane solutions in type~0B gravity theory.
Indeed, a $7$-brane solution in type~IIB supergravity is also a solution in type~0B gravity theory under the $Q$-symmetric ansatz.
Hence, under the $Q$-symmetric ansatz, we can construct $7$-brane solutions in type~0B gravity theory in the same way as the BPS $7$-brane solutions of type~IIB supergravity are constructed.
In other words, the holomorphic ansatz still solves the equation of motion for the axio-dilaton field $\tau$ under the $Q$-symmetric ansatz, although type~0B gravity theory does not possess spacetime supersymmetry in contrast to type~IIB supergravity.

More specifically, the D$7$-brane solution described by Eq.~\eqref{eq:D7_tau_sol} and Eq.~\eqref{eq:D7_metric_sol} still exists in type~0B gravity theory.
Since the monodromy of the D$7$-brane in this solution is $\sT\in\Gamma_{0}(2)$, it carries a unit charge under the operator $\cQ^{\bZ}_{\gamma}$ and no charge under $\cQ^{\bZ_{4}}_{\gamma}$.

As observed above, the equations of motion for the metric and the first RR $0$-form field $C^{(0)}$ under the $Q$-symmetric ansatz coincide with those of type~IIB supergravity.
This suggests that the F-theoretic backgrounds of type~IIB supergravity are also solutions of type~0B gravity theory.
It is necessary to take into account two differences between type~IIB supergravity and type~0B gravity theory.
First, any element of $\SL(2,\bZ)$ is allowed as the monodromy in type~IIB supergravity, while only the elements of $\Gamma_{0}(2)$ are allowed in type~0B gravity theory under the $Q$-symmetric ansatz.
Second, the equations of motion for the small fluctuations of the $Q$-odd fields must be well-defined in a consistent background of type~0B gravity theory.
Let us explain why we impose the second condition.
By a background or a solution, we mean a saddle point of the path integral, around which the fluctuations of all the fields of the theory can be defined.
In an F-theoretic background, the axio-dilaton field $\tau$ is defined only up to the monodromy, and the fields are glued to each other across the branch cuts by the corresponding duality transformations.
Such a gluing makes sense only if the duality group acts on all the fields of the theory, and not merely on those retained by the truncation.
In type~IIB supergravity, this requirement is automatically fulfilled, since $\SL(2,\bZ)$ is a symmetry of the full theory and the transformation law of every field is known.
On the other hand, in type~0B gravity theory, the $\Gamma_{0}(2)$ symmetry has been argued for only in the $Q$-symmetric branch, and the transformation law of the $Q$-odd fields is not known.
Hence, the consistency of the gluing for the $Q$-odd fields imposes a nontrivial condition on the background.
In what follows, we examine the constraints that these two conditions impose on the background.

It is a classical fact that any elliptically fibered K3 surface whose monodromies take values in $\Gamma_{0}(2)$ can be written in the form
\begin{align}\label{eq:restricted_Weierstrass}
	y^{2}=x\left(x^{2}+a(z)x+b(z)\right),
\end{align}
where $a(z)$ and $b(z)$ are polynomials of degree four and eight, respectively.\footnote{See, e.g.,~\cite{Bershadsky:1998vn} for a discussion of the restricted Weierstrass form~\eqref{eq:restricted_Weierstrass} in the context of string theory.}
For the above Weierstrass form~\eqref{eq:restricted_Weierstrass}, the discriminant and the $j$-function are given by
\begin{align}\begin{aligned}\label{eq:invariants}
	\Delta_{W}&=16b^{2}D,&
	D&=a^{2}-4b,&
	j&=\frac{16^{3}(a^{2}-3b)^{3}}{\Delta_{W}}.
\end{aligned}\end{align}
In the present situation, the modular functions~\eqref{eq:modular_func_0B} take the form
\begin{align}\begin{aligned}
\Delta^{\U(1)}&=\frac{D}{2^{8}b},&
\Delta^{\bZ_{4}}&=\frac{(2\pi)^{4}}{\omega_{1}^{4}b},
\end{aligned}\end{align}
where $\omega_{1}$ is defined in Appendix~\ref{subsec:lambda}.
It follows from these expressions that the charge operator~\eqref{eq:charge_op_0B} reduces to
\begin{align}
	\cQ^{\bZ}_\gamma
	=\sum_{z_*:\,\text{locus inside\,}\gamma}\left(
	\ord_{z=z_{\ast}} D(z)-\ord_{z=z_{\ast}} b(z)
	\right).
\end{align}
Note that $\Delta^{\bZ_{4}}$ is not a single-valued function, since a generic element of $\Gamma_{0}(2)$ transforms $\omega_{1}$ as
\begin{align}\begin{aligned}
	\sM\cdot\omega_{1}
	&=(r\tau+s)\omega_{1},&
	\sM&=\begin{pmatrix}
	p&q\\
	r&s
	\end{pmatrix}\in\Gamma_{0}(2).
\end{aligned}\end{align}
This factor $(r\tau+s)$ is precisely compensated for by the term $F$ in the definition~\eqref{eq:precise_Z4_charge} of the charge operator.
We then obtain
\begin{align}
	\cQ^{\bZ_{4}}_\gamma=-\sum_{z_*:\,\text{locus inside\,}\gamma} \ord_{z=z_{\ast}} b(z).
\end{align}

For generic $a$ and $b$, the discriminant has eight simple zeros on the locus $D=0$ and eight double zeros on the locus $b=0$.
In the Kodaira classification, they correspond to eight $I_{1}$ fibers and eight $I_{2}$ fibers, respectively.
It should be emphasized that the two loci are distinguished not by their Kodaira type but by the $\Gamma_{0}(2)$ conjugacy class of the monodromy.
As explained in Appendix~\ref{subsec:Monodromy_Charge}, the monodromy of the $I_{1}$ fibers on the locus $D=0$ is conjugate in $\Gamma_{0}(2)$ to $\sT$, whereas that of the $I_{2}$ fibers on the locus $b=0$ is conjugate to $(\sX\sT)^{-1}$.
Their charges are therefore given by
\begin{align}\label{eq:generic_charges}
	\left(\cQ^{\bZ}_{\gamma}(\sT),\cQ^{\bZ_{4}}_{\gamma}(\sT)\right)=(1,0),
	\qquad
	\left(\cQ^{\bZ}_{\gamma}(\sT^{-1}\sX^{-1}),\cQ^{\bZ_{4}}_{\gamma}(\sT^{-1}\sX^{-1})\right)=(-1,-1).
\end{align}
In particular, the total charge of the generic configuration indeed vanishes, i.e.~$8\,(1,0)+8\,(-1,-1)=(0,-8)\equiv(0,0)$.

We now turn to a configuration in which each $7$-brane carries a unit of $\bZ_{4}$ charge.
It is obtained by setting $a=0$ in~\eqref{eq:restricted_Weierstrass},
\begin{align}\label{eq:Z4_model}
	y^{2}=x^{3}+b(z)x,
\end{align}
where $b(z)$ is a generic polynomial of degree eight.
Then, we have
\begin{align}\begin{aligned}
	D&=-4b,&
	\Delta_{W}&=-64b^{3},&
	j&=1728.
\end{aligned}\end{align}
All eight singular fibers correspond to Kodaira type $III$, whose monodromy is conjugate to $\sX^{-1}$ in $\Gamma_{0}(2)$ as discussed in Appendix~\ref{subsec:Monodromy_Charge}.
The charges of these $7$-branes are given by
\begin{align}
	\left(\cQ^{\bZ}_{\gamma}(\sX^{-1}),\cQ^{\bZ_{4}}_{\gamma}(\sX^{-1})\right)=(0,-1).
\end{align}
Again, the total charge is trivial as an element of $\bZ\times\bZ_{4}$.
This is the explicit realization of the second type of $7$-brane predicted by the Abelianization.
Since the $j$-function is constant in this background, the axio-dilaton does not vary over the $z$-plane.
Because a constant $\tau$ must be invariant under $\sX^{-1}$, its value is fixed to be
\begin{align}\label{eq:fixed_tau_X}
	\tau=\frac{1+\i}{2}.
\end{align}
This configuration is the type~0B counterpart of the F-theory background discussed in Refs.~\cite{Sen:1996vd,Dasgupta:1996ij}.

Next, we discuss the linearized equations of motion for small fluctuations of the $Q$-odd fields around these backgrounds.
From the action~\eqref{eq:axio_dilaton_frame}, we obtain them as
\begin{align}\begin{aligned}\label{eq:Q-odd_eom}
	&\Box_{8}T+4e^{-\rho}\del_{z}\del_{\zbar}T-\meff^{2}\tau_{2}^{-1/2}T=\frac{1}{2\sqrt{2}}e^{-\rho}\tau_{2}^{-2}\left(\del_{z}\tau\del_{\zbar}\tilde{C}^{(0)}+\del_{\zbar}\taubar\del_{z}\tilde{C}^{(0)}\right),\\
	&\meff^{2}=-\frac{2}{\slope}+\frac{\left|\del_{z}\tau\right|^{2}}{4\,\tau_{2}^{3/2}\,e^{\rho}},\\
	&\Box_{8}\tilde{C}^{(0)}+2e^{-\rho}\tau_{2}^{2}\left(\del_{z}(\tau_{2}^{-2}\del_{\zbar}\tilde{C}^{(0)})+\del_{\zbar}(\tau_{2}^{-2}\del_{z}\tilde{C}^{(0)})\right)=-\frac{1}{2\sqrt{2}}e^{-\rho}\left(\del_{z}\tau\del_{\zbar}T+\del_{\zbar}\taubar\del_{z}T\right),
\end{aligned}\end{align}
where $\Box_{8}$ is the d'Alembertian of the flat $(1+7)$-dimensional Minkowski space.
In the present setup, the axio-dilaton field is allowed to have a monodromy valued in $\Gamma_{0}(2)$.
As a result, the equations~\eqref{eq:Q-odd_eom} are ill-defined in the background described by the restricted Weierstrass form~\eqref{eq:restricted_Weierstrass} with generic $a$ and $b$: assuming that the $Q$-odd fields are neutral under $\Gamma_{0}(2)$, the equations are not invariant under a general monodromy transformation in $\Gamma_{0}(2)$.\footnote{We do not rule out the possibility that the equations of motion could become globally well-defined for some highly non-trivial assignment of $\Gamma_{0}(2)$ transformations to the $Q$-odd fields. See Ref.\cite{Baykara:2026gem} for a related discussion.}

The situation is particularly simple for the background described by the model~\eqref{eq:Z4_model}.
In this background, since the axio-dilaton field is constant, the equations of motion~\eqref{eq:Q-odd_eom} reduce to
\begin{align}\begin{aligned}
	\left(\Box_{8}+4e^{-\rho}\del_{z}\del_{\zbar}-\frac{2}{\slope}\tau_{2}^{-1/2}\right)T&=0,&
	\left(\Box_{8}+4e^{-\rho}\del_{z}\del_{\zbar}\right)\tilde{C}^{(0)}&=0.
\end{aligned}\end{align}
These equations are manifestly globally well-defined on the base.
Because $\tau$ does not vary, the RR $0$-form field does not contribute to the effective mass, and the coefficient in the equation of motion for $T$ is the constant $-2/(\slope\sqrt{\tau_{2}})$ with $\tau_{2}=1/2$.
As the base space is compact, the lowest Kaluza--Klein mode is the constant mode, whose eight-dimensional mass squared is exactly this constant.
We emphasize that this tachyonic mode is precisely the one already present in the flat ten-dimensional vacuum of type~0B string theory, and that the $7$-brane background described by the model~\eqref{eq:Z4_model} neither removes this instability nor enhances it.
Note also that this conclusion is independent of the overall size of the base space, since the effective mass is constant over the base.
In this sense, the $7$-branes with the charges $(0,-1)\in\bZ\times\bZ_{4}$ do not lift the closed string tachyon.

A second background in which the equations~\eqref{eq:Q-odd_eom} are well-defined is the local solution given by~\eqref{eq:D7_tau_sol} and~\eqref{eq:D7_metric_sol}.
The monodromy of the $7$-brane in this solution is $\sT\in\Gamma_{0}(2)$.
One can easily check that the equations of motion~\eqref{eq:Q-odd_eom} are invariant under the monodromy $\sT$ and hence globally well-defined.
In contrast to the model~\eqref{eq:Z4_model}, the axio-dilaton field varies over the $z$-plane in this background, and hence the kinetic energy of the first RR $0$-form field contributes to the effective mass of the tachyon field.
Near the $7$-brane, the logarithmic solution~\eqref{eq:D7_tau_sol} gives $\tau_{2}\sim\frac{1}{2\pi}\log(1/|z|)\to\infty$, and the metric~\eqref{eq:D7_metric_sol} behaves as $e^{\rho}\propto\tau_{2}$.
Therefore, the coefficient in the equation of motion for $T$ behaves as
\begin{align}
	\meff^{2}\tau_{2}^{-1/2}\ \sim\ -\frac{2}{\slope\sqrt{\tau_{2}}}+\frac{c}{|z|^{2}\tau_{2}^{3}},
\end{align}
with a positive constant $c$.
The two terms work in the same direction as one approaches the $7$-brane: the string coupling becomes weak, so that the contribution of the bare tachyon mass is suppressed, while the contribution of the RR $0$-form field grows.
The coefficient is therefore expected to change sign at some radius.
The solution~\eqref{eq:D7_tau_sol} and~\eqref{eq:D7_metric_sol} is a local description valid only in the vicinity of the $7$-brane.
In particular, $\tau_{2}$ decreases as $|z|$ grows, and the solution breaks down at large $|z|$.
The behavior of the tachyon field away from the $7$-brane is thus not determined by this solution, but by a global completion of the background, which is not provided here.
Note that the base space cannot be compactified using only $7$-branes of this type, since such a configuration is inconsistent with the restricted Weierstrass form~\eqref{eq:restricted_Weierstrass}.

Let us summarize the consequences of the two conditions.
Among the backgrounds discussed above, the model~\eqref{eq:Z4_model} and the local solution~\eqref{eq:D7_tau_sol} and~\eqref{eq:D7_metric_sol} satisfy both of them, whereas the background described by the restricted Weierstrass form~\eqref{eq:restricted_Weierstrass} with generic $a$ and $b$ does not satisfy the second one.
These two backgrounds contain $7$-branes whose monodromies are conjugate to $\sX^{-1}$ and to $\sT$, that is, precisely the generators of the $\bZ_{4}$ factor and of the $\bZ$ factor of $\Gamma_{0}(2)^{\rm ab}$.
Therefore, the charged objects of both charge operators are realized in backgrounds of type~0B gravity theory in which the equations of motion for the $Q$-odd fields are globally well-defined.

We close this section with a comment on the status of the background with generic $a$ and $b$.
The charges assigned in~\eqref{eq:generic_charges} are determined by the monodromy alone, which is well-defined for any solution satisfying the $Q$-symmetric ansatz.
They are therefore unaffected by the second condition, and the same remark applies to the identification of the charges in Appendix~\ref{subsec:Monodromy_Charge}, where a deformation of the model~\eqref{eq:Z4_model} into a generic configuration is used.
The only question left open is whether such a background can be promoted to a background of the full type~0B string theory, which would require knowledge of the $\Gamma_{0}(2)$ transformation law of the $Q$-odd fields.

\section{Discussions}
\label{sec:Discussions}

In Section~\ref{subsec:0B_7brane_sol}, we studied the model~\eqref{eq:Z4_model}, in which the axio-dilaton field takes the constant value~\eqref{eq:fixed_tau_X}.
It is worth comparing this value with the result of Ref.~\cite{Baykara:2026gem}, where the potential of type~0B string theory is analyzed on the $Q$-symmetric branch and regarded as a function of the axio-dilaton field.
The critical point of this potential is found to lie exactly at the value~\eqref{eq:fixed_tau_X}, and the corresponding solution is identified as an unstable de Sitter background.
We emphasize that no potential for the axio-dilaton field is present in the truncated action~\eqref{eq:axio_dilaton_frame}, which coincides with that of type~IIB supergravity.
The instability found in Ref.~\cite{Baykara:2026gem}, which lies in the direction of the axio-dilaton field, is accordingly invisible in our analysis.
What we found instead is an instability in the $Q$-odd sector.
Around this background the linearized equations of motion for the $Q$-odd fields are well-defined, and the effective mass of the tachyon field receives no contribution from the RR $0$-form field, so that the closed string tachyon survives.

In deriving the second condition imposed in Section~\ref{subsec:0B_7brane_sol}, we assumed that the $Q$-odd fields are neutral under $\Gamma_{0}(2)$, and it was precisely this assumption that rendered the linearized equations~\eqref{eq:Q-odd_eom} ill-defined in the background with generic $a$ and $b$.
This assumption need not be the correct one.
A proposal for the structure of the $Q$-odd sector has been put forward in Ref.~\cite{Baykara:2026gem}.
If the duality transformation of the $Q$-odd fields can be extracted from this proposal, one could revisit the second condition: the terms in~\eqref{eq:Q-odd_eom} that fail to be single-valued may combine into a covariant expression, in which case the F-theoretic backgrounds with generic $a$ and $b$ would be promoted to genuine backgrounds of the full type~0B string theory.
This would considerably enlarge the class of $7$-brane backgrounds available in type~0B string theory.

It would be interesting to determine the degrees of freedom living on a $7$-brane carrying $\bZ_{4}$ charge.
In conventional F-theory, the gauge algebra supported on a $7$-brane is read off from the Kodaira type of the corresponding singular fiber.
In the present case, however, the Kodaira type alone does not characterize the $7$-brane, since two $7$-branes of the same Kodaira type may belong to different $\Gamma_{0}(2)$ conjugacy classes and carry different charges.
One may therefore expect the gauge degrees of freedom on a type~0B $7$-brane to be labeled by the $\Gamma_{0}(2)$ conjugacy class of its monodromy.

In this paper, we have focused on the $\Gamma_{0}(2)$ duality of type~0B string theory, which originates from M-theory compactified on a torus with periodic and antiperiodic boundary conditions along the two cycles.
On the other hand, it has been proposed in Ref.~\cite{Baykara:2026gem} that type~0B string theory is dual to M-theory on $S^{1}\times(S^{1}\vee S^{1})$, and this proposal suggests that the duality group on the $Q$-symmetric branch is instead $\Gamma^{0}(2)$.
Since $\Gamma^{0}(2)=\sS\,\Gamma_{0}(2)\,\sS^{-1}$, the two groups are conjugate in $\SL(2,\bZ)$, and their Abelianizations are isomorphic.
It would therefore be interesting to construct the charge operators associated with the Abelianization of $\Gamma^{0}(2)$ and to identify the corresponding $7$-branes.

\section*{Acknowledgements}
The authors would like to thank Arata Ishige, Shotaro Kawanago, and Shun'ya Mizoguchi for their helpful discussions.
The work of NK is supported by 
JSPS KAKENHI Grant Number 
JP24KJ0157.
The work of MK is supported by 
JSPS KAKENHI Grant Numbers 
25K23381, 25K01002, 
and 
the COREnet project (COREnet062) of
Research Center for Nuclear Physics, 
Osaka University.
The work of HW is supported in part by JST FOREST Program (Grant Number JPMJFR2030, Japan).

\appendix

\section{Relation between Charges of 7-branes and Weierstrass Form}
\label{sec:Charge_Weierstrass}

In this appendix, we determine the charges of $7$-branes described by the Weierstrass form.
The results in this appendix are used in Section~\ref{subsec:0B_7brane_sol} to identify the charges of $7$-branes in the gravity solutions.

\subsection{Modular \texorpdfstring{$\lambda$}{lambda} Function and Roots of Cubic Equation}
\label{subsec:lambda}

Since the right-hand side of the Weierstrass form~\eqref{eq:Weierstrass} is a cubic polynomial in $x$, we can always express it in terms of three roots $r_{1}$, $r_{2}$, $r_{3}$ as
\begin{align}
	y^{2}=(x-r_{1})(x-r_{2})(x-r_{3}).
\end{align}
For the restricted Weierstrass form~\eqref{eq:restricted_Weierstrass}, we have the following relations,
\begin{align}
	r_{1}&=0,&
	r_{2}+r_{3}&=-a,&
	r_{2}r_{3}&=b,&
	D=a^{2}-4b&=(r_{2}-r_{3})^{2}.
\end{align}
Since $(x,y)=(0,0)$ is a globally defined section over the base, the root $r_{1}=0$ is distinguished.
In contrast, the two roots $r_{2,3}=\tfrac12(-a\pm\sqrt{D})$ of the quadratic factor may be exchanged when $z$ is transported along a closed loop.

The complex structure $\tau$ of the fiber is encoded in the roots.
It is well known that one can order the roots as $(e_{1},e_{2},e_{3})$ such that
\begin{align}\begin{aligned}
	e_{1}-e_{2}&=\left(\frac{2\pi}{\omega_{1}}\right)^{2}\vartheta_{3}^{4},&
	e_{1}-e_{3}&=\left(\frac{2\pi}{\omega_{1}}\right)^{2}\vartheta_{4}^{4},&
	e_{3}-e_{2}&=\left(\frac{2\pi}{\omega_{1}}\right)^{2}\vartheta_{2}^{4},
\end{aligned}\end{align}
where $\omega_{1}$ is the period of the holomorphic $1$-form along the $A$-cycle of the elliptic curve,
\begin{align}\begin{aligned}
	\omega_{1}=\oint_{A}\frac{\d x}{y}.
\end{aligned}\end{align}

Throughout this appendix we adopt the convention that the distinguished root is assigned to the first slot, $e_{1}=r_{1}=0$, while the two undistinguished roots $r_{2}$ and $r_{3}$ are assigned to $e_{2}$ and $e_{3}$ in either order.
With this convention, we have
\begin{align}\begin{aligned}\label{eq:b_D_theta}
	b&=(e_{1}-e_{2})(e_{1}-e_{3})=\left(\frac{2\pi}{\omega_{1}}\right)^{4}\vartheta_{3}^{4}\vartheta_{4}^{4},&
	D&=(e_{3}-e_{2})^{2}=\left(\frac{2\pi}{\omega_{1}}\right)^{4}\vartheta_{2}^{8}.
\end{aligned}\end{align}
Combining these expressions with the identities
\begin{align}\label{eq:theta_eta_identities}
	\vartheta_{2}(\tau)^{4}=\frac{16\,\eta(2\tau)^{8}}{\eta(\tau)^{4}},
	\qquad
	\vartheta_{2}(\tau)\vartheta_{3}(\tau)\vartheta_{4}(\tau)=2\,\eta(\tau)^{3},
\end{align}
we obtain the expressions for the two modular functions~\eqref{eq:modular_func_0B} in terms of the Weierstrass form as
\begin{align}\begin{aligned}\label{eq:modular_func_Weierstrass}
	\Delta^{\U(1)}(\tau)=\frac{\eta(2\tau)^{24}}{\eta(\tau)^{24}}
	&=\frac{\vartheta_{2}^{8}}{2^{8}\,\vartheta_{3}^{4}\vartheta_{4}^{4}}
	=\frac{D}{2^{8}\,b},\\
	\Delta^{\bZ_{4}}(\tau)=\frac{\eta(2\tau)^{8}}{\eta(\tau)^{16}}
	&=\frac{1}{\vartheta_{3}^{4}\vartheta_{4}^{4}}
	=\left(\frac{2\pi}{\omega_{1}}\right)^{4}\frac{1}{b},
\end{aligned}\end{align}
which are the expressions used in Section~\ref{subsec:0B_7brane_sol}.
Note that $\Delta^{\U(1)}$ is a ratio of two polynomials of the same degree and is therefore a genuine function on the base, in accordance with the fact that it is a modular function of weight zero.
By contrast, $\Delta^{\bZ_{4}}$ has weight $-4$, and correspondingly its Weierstrass expression involves the period $\omega_{1}$, which is not single-valued on the base.

The modular $\lambda$-function is defined as
\begin{align}\label{eq:roots_lambda}
	\lambda(\tau)&=\frac{e_{3}-e_{2}}{e_{1}-e_{2}}=\frac{\vartheta_{2}(\tau)^{4}}{\vartheta_{3}(\tau)^{4}}
	=16\,e^{\i\pi\tau}\left(1+\cO\!\left(e^{\i\pi\tau}\right)\right).
\end{align}
It follows from the modular properties of the Jacobi $\vartheta$-functions that
\begin{align}\label{eq:lambda_modular}
	\lambda(\tau+1)&=\frac{\lambda}{\lambda-1},&
	&\lambda(-1/\tau)=1-\lambda.
\end{align}
From these relations, one can check that $\lambda(\sM\tau)=\lambda(\tau)$ for $\sM\in\Gamma(2)$, which means that the group $\SL(2,\bZ)/\Gamma(2)$ acts on the $\lambda$-function.
On the other hand, the permutations of $(e_{1},e_{2},e_{3})$ also act on $\lambda$, and these actions match those generated by the modular transformations~\eqref{eq:lambda_modular}.
In particular, the exchange of $e_{2}$ and $e_{3}$ yields
\begin{align}\label{eq:roots_swap}
	\frac{e_{2}-e_{3}}{e_{1}-e_{3}}
	=\frac{\lambda}{\lambda-1},
\end{align}
i.e.,~the exchange of the two undistinguished roots corresponds to the monodromy $\sT$.

Along a closed loop in the $z$-plane, the roots may be permuted.
Since $r_{1}=0$ is globally defined for the restricted Weierstrass form~\eqref{eq:restricted_Weierstrass}, the permutation fixes $e_{1}$ and can at most exchange $e_{2}$ and $e_{3}$.
From the observation~\eqref{eq:roots_swap}, the monodromy is therefore the identity or $\sT$ up to $\Gamma(2)$.
Since $\Gamma_{0}(2)/\Gamma(2)$ is generated by $\sT$, the monodromy must belong to $\Gamma_{0}(2)$.

It follows from Eq.~\eqref{eq:roots_lambda} that when $e_{2}$ and $e_{3}$ collide, $\lambda$ approaches zero and $\tau\to\i\infty$.
On the other hand, $\tau\to 0$ corresponds to the collision of $e_{1}$ and $e_{2}$.
The $\sT$-transformation maps the collision of $e_{1}$ and $e_{2}$ to that of $e_{1}$ and $e_{3}$.

\subsection{Monodromy and Charges of 7-branes}
\label{subsec:Monodromy_Charge}
In Section~\ref{subsec:0B_7brane_sol}, we considered the generic configurations described by the restricted Weierstrass form~\eqref{eq:restricted_Weierstrass}.
The discriminant is given by $\Delta_{W}=16b^{2}D$, where $D=a^{2}-4b$ is the discriminant of the quadratic factor, whose roots are the undistinguished ones.
For generic $a(z)$ and $b(z)$, eight $I_{1}$ fibers appear from simple zeros on the locus $D=0$, whereas eight $I_{2}$ fibers arise from double zeros on the locus $b=0$.
Hence, we discuss the monodromy and charges of $7$-branes on $D=0$ and $b=0$ in turn.

We begin with a $7$-brane on the locus $D=0$.
Without loss of generality, we may assume that the $7$-brane is located at $z=0$, and we further assume $\ord_{z=0}D=1$ and $\ord_{z=0}a=\ord_{z=0}b=0$.
Then, the $\lambda$-function is given by
\begin{align}
	\lambda=\frac{r_{3}-r_{2}}{-r_{2}}
	=\frac{\sqrt{D}}{-r_{2}}\ \propto\ z^{1/2}.
\end{align}
By comparing this result with Eq.~\eqref{eq:roots_lambda}, we obtain $e^{\i\pi\tau}\propto z^{1/2}$, which means
\begin{align}
	\tau(z)=\frac{1}{2\pi\i}\log z+(\text{regular}).
\end{align}
As expected, the monodromy along the loop around $z=0$ is $\sM=\sT\in\Gamma_{0}(2)$.
Thus, the charges of this $7$-brane are given by
\begin{align}
	\left(\cQ^{\bZ}_{\gamma}(\sT),\cQ^{\bZ_{4}}_{\gamma}(\sT)\right)=\,(1,0).
\end{align}

Next, we discuss the $7$-brane on the locus $b=0$.
To this end, we assume $\ord_{z=0}b=1$ and $\ord_{z=0}a=0$.
Now $r_{2}=-\,b(1+\cO(b))/a\propto z$ collides with the distinguished root $r_{1}=0$, while $r_{3}=-a+\cO(b)$ stays away from the origin.
By assigning the colliding root to the third slot, $e_{2}=r_{3}$ and $e_{3}=r_{2}$, we have
\begin{align}
	\lambda=\frac{e_{3}-e_{2}}{e_{1}-e_{2}}
	=\frac{-b/a+a}{a}
	=1-\frac{b}{a^{2}}+\cO(b^{2}).
\end{align}
Hence, $\lambda\to1$ as $z$ approaches zero.
As mentioned above, this behavior implies that $\tau$ approaches the cusp $\tau=0$.
To read off the monodromy, we work with the coordinate $w$ adapted to this cusp as
\begin{align}
	\tau=-\frac{1}{w}=\sS\cdot w,\qquad w\to\i\infty,
\end{align}
for which $\lambda(w)=1-\lambda(\tau)\propto z$ by Eq.~\eqref{eq:lambda_modular}.
By comparing this result with the expression $\lambda(w)=16e^{\i\pi w}(1+\cO(e^{\i\pi w}))$, we find that $e^{\i\pi w}\propto z$.
Thus, the monodromy along the closed loop is $\sT^{2}$ in the $w$ frame.
Since $\tau=\sS w$, the monodromy in the original frame is obtained as
\begin{align}
	\sS\sT^{2}\sS^{-1}=\begin{pmatrix}1&0\\-2&1\end{pmatrix}=(\sX\sT)^{-1}.
\end{align}
Then, the charges of the corresponding $7$-brane are given by
\begin{align}
	\left(\cQ^{\bZ}_{\gamma}(\sT^{-1}\sX^{-1}),\cQ^{\bZ_{4}}_{\gamma}(\sT^{-1}\sX^{-1})\right)=(-1,-1).
\end{align}

Finally, we determine the charges of the $7$-brane in the model~\eqref{eq:Z4_model}, in which $a$ vanishes identically.
Again, we assume $\ord_{z=0}b=1$.
The analysis of the preceding paragraph cannot be applied directly, since it requires $\ord_{z=0}a=0$.
We therefore deform the Weierstrass form by an infinitesimal constant $\epsilon$ as
\begin{align}
	y^{2}=x^{3}+\epsilon x^{2}+bx,
\end{align}
so that $a=\epsilon$ and hence $\ord_{z=0}a=0$; the results obtained above then become available.
The discriminant $\Delta_{W}=16b^{2}D$ now vanishes at two nearby points.
At $z=0$, the polynomial $b$ still has a simple zero.
As derived in the previous paragraph, the charges of this $7$-brane are given by $(-1,-1)\in\bZ\times\bZ_{4}$.
On the other hand, at $z=\epsilon^{2}/\left(4b'(0)\right)+\cO(\epsilon^{3})$, the discriminant $D=\epsilon^{2}-4b$ has a simple zero.
The corresponding $7$-brane has the charges $(1,0)\in\bZ\times\bZ_{4}$.
In the limit $\epsilon\to0$, the two $7$-branes merge.
Hence, the charge of the resulting $7$-brane is given by
\begin{align}\label{eq:III_charge}
	(1,0)+(-1,-1)=(0,-1)\in\bZ\times\bZ_{4}.
\end{align}
This charge coincides with that of $\sX^{-1}$.

\bibliographystyle{ytphys}
\bibliography{main}
\end{document}